\documentclass[sort&compress,preprint]{aipproc}
\usepackage{graphicx}
\usepackage{amsmath}
\layoutstyle{8x11single}
\newcommand{\tens}[1]{\ensuremath{{\bf #1}}}

\newcommand{\id}{\tens{1}}
\newcommand{\I}{\mathrm{i}}

\renewcommand{\Re}{\ensuremath{\operatorname{Re}}}
\renewcommand{\Im}{\ensuremath{\operatorname{Im}}}

\newcommand{\sign}{\ensuremath{\operatorname{sgn}}}

\newcommand{\emme}[1]{\tens{\bf M}_{#1}}
\newcommand{\enne}[1]{\tens{\bf N}_{#1}}

\newcommand{\evmax}{\Delta\epsilon}

\newcommand{\q}{l}

\begin{document}

\title{A hybrid approach to Fermi operator expansion}
\newcommand{\ethzaffiliation}{{Computational Science, Department of Chemistry and Applied Biosciences,
ETH Zurich, USI Campus, Via Giuseppe Buffi 13, CH-6900 Lugano, Switzerland}}
\author{Michele Ceriotti}{%
address=\ethzaffiliation,
,email={michele.ceriotti@phys.chem.ethz.ch}}
\author{Thomas D. K\"uhne}{address=\ethzaffiliation}
\author{Michele Parrinello}{address=\ethzaffiliation}

\begin{abstract}
In a recent paper we have suggested that the finite temperature density matrix
can be computed efficiently by a combination of polynomial expansion and 
iterative inversion techniques. We present here significant improvements 
over this scheme. The original complex-valued formalism is turned into a purely 
real one. In addition, we use Chebyshev polynomials expansion and fast 
summation techniques. This drastically reduces the scaling of the algorithm
with the width of the Hamiltonian spectrum, which is now of the order of the
cubic root of such parameter.
This makes our  method very competitive for applications to {\em ab-initio} 
simulations, when high energy resolution is required.
\end{abstract}


\maketitle

\section{Introduction}
Several fields of computational science (nanotechnology, materials science or biochemistry 
just to name a few) would greatly benefit from the possibility of performing simulations
of large systems, containing hundreds of thousands of atoms.
Conventional electronic structure calculations require the diagonalization
of matrices whose size $N$ is of the order of the number of electrons in the system.
The $\mathcal{O}\left(N^3\right)$-scaling cost of this step greatly limits the 
range of systems which can be tackled by {\em ab-initio} techniques, despite the fast-paced progress in 
the computational power of modern processors.
Based on the theoretical foundations of the nearsightedness principle of electronic 
matter\cite{kohn96prl}, several techniques have been developed in the last years to avoid
the diagonalization step, by directly computing the density matrix of the system using linear
scaling algorithms\cite{goed99rmp,yang91prl,gall-parr92prl,li+93prb,baro-gian92epl,pals-mano98prb}.
One of the earliest approaches have been to compute the finite temperature 
density matrix, i.e. the Fermi function of the system's Hamiltonian, $\tens{H}$, by decomposing it 
into easier-to-compute functions of the Hamiltonian\cite{goed-colo94prl,goed-tete95prb},
for instance Chebyshev polynomials or rational functions.
In a recent paper\cite{ceri+08jcp} we have discussed an exact decomposition of the Fermi operator 
which can be efficiently computed by a combination of polynomial expansion
and iterative inversion techniques. In this way, we achieved an efficient scaling with $\evmax$,  
the width of the spectrum of the Hamiltonian in units of $k_BT$.
This makes the method attractive for applications to metals or low-band gap semiconductors
at low electronic temperature.
In this short paper we discuss a number of improvements to this scheme, which further
lower the operation count, leading to a scaling $\propto \sqrt[3]{\Delta\epsilon}$, 
which is, to the best of our knowledge, the lowest so far reported in literature.
We will follow closely the scheme of our previous work\cite{ceri+08jcp}, obtaining analytical
estimates for the operation count of the different steps which compose our algorithm, so as to 
optimize them in order to achieve optimal performance.

\section{Details of the decomposition}
Our decomposition scheme is based on an exact expansion of the Fermi operator,
which can be elegantly derived using the grand-canonical formalism\cite{alav+94prl}. 
Here we only report the final result, namely the fact that the finite-temperature density matrix $\tens{\boldsymbol{\rho}}$ 
can be written in terms of a sum of complex-valued matrices $\emme{\q}$\cite{kraj-parr05prb}: 
\begin{equation}
\tens{\boldsymbol{\rho}}=\frac{\delta\Omega}{\delta\tens{H}}=
\frac{1}{1+e^{\tens{H}}}=
\frac{2}{P}\sum_{\q=1}^{P}\id-\Re\emme{\q}^{-1},\quad
\emme{\q}=\id-e^{\I\left(2\q-1\right)\pi/2P}e^{-\tens{H}/2P}. \label{eq:expansion}
\end{equation}
Equation~(\ref{eq:expansion}) is exact for any value of $P$, but large values should
be chosen so as to allow for simple and economical evaluation of the matrix exponential 
$e^{-\tens{H}/2P}$. Here and in the rest of the paper, with no loss of generality, we use a shifted and scaled 
Hamiltonian, i.e. we set the zero of energies at $\mu$, and measure energy in units of $k_B T$.

The expensive step in applying Eq.~(\ref{eq:expansion}) is the inversion of the $\emme{\q}$
matrices. In Ref.\cite{ceri+08jcp} we have shown that, for large values of $P$, a $\bar{\q}$ exists such that
all of the matrices with $\q>\bar{\q}$ are almost optimally conditioned, and can be easily inverted 
 by a polynomial expansion. We have also shown how their overall contribution to the Fermi operator 
can be computed at once, at the same cost of computing a single term, making the 
computational burden of the method virtually independent of $P$. We will exploit this property and 
take often the large $P$ limit to derive analytical results.
The remaining, low-$\q$ matrices are more efficiently treated
with an iterative Newton inversion scheme\cite{pan-reif85proc}. 
Overall, the scaling in terms of matrix-matrix multiplications 
count, which is $\propto \evmax$ for standard Chebyshev polynomials expansion\cite{baer-head97jcp},
reduces to $\propto \sqrt{\evmax}$, and is even lower when fast polynomial summation 
methods\cite{vanl79ieee,lian+03jcp} are used.

A first improvement over the initial formulation of our scheme can be seeked by noting
that only the real part of $\emme{\q}^{-1}$ enters the expression for the Fermi operator,
so that it can be more efficient to write 
\begin{equation}
\Re \emme{\q}^{-1}=\frac{1}{2}\left[\id +\left(e^{\tens{H}/P}-\id\right)\enne{\q}^{-1}\right],\text{ where }
\quad \enne{\q}=\id + e^{\tens{H}/P}-2 e^{ \tens{H}/2P}\cos \pi\frac{2\q-1}{2P}.
\label{eq:def-enne}
\end{equation}
Only real matrices are involved in Eq.~(\ref{eq:def-enne}), leading to substantial savings
in memory requirements and computation time. The inversion of the $\enne{\q}$s is the computationally
demanding part of this approach; in fact, if one performs an analysis similar to the one 
we have carried out in Ref.\cite{ceri+08jcp}, the condition number of $\enne{\q}$ is found to be
approximately $\kappa\left(\enne{\q}\right)\approx 1+\evmax^2 \pi^{-2} \left(2\q-1\right)^{-2}$.
This is higher than  $\kappa\left(\emme{\q}\right)\approx 1+\evmax \pi^{-1} \left(2\q-1\right)^{-1}$, but
decreases more rapidly with $\q$.
The low-$\q$ inverse matrices, which are worse conditioned, can  be computed by Newton inversion, 
whose performances are only weakly affected by high condition numbers. 

Moreover, we take profit of the form of Eq.~(\ref{eq:def-enne}) to improve the evaluation
of the contribution of the high-$\q$ matrices, which in our previous work\cite{ceri+08jcp} had a computational cost 
scaling as $\evmax^2/\bar{\q}^2$. Since we also want to set up an expansion in Chebyshev polynomials
for $\enne{\q}^{-1}$, we rewrite $\enne{\q}$ it in terms of an auxiliary matrix $\tens{X}$ whose
spectrum lies between $-1$ and $1$:
\begin{equation}
\enne{\q}=\zeta^2 \tens{X}^2-2\zeta\left(\cos\pi\frac{2\q-1}{2P}-z_0\right)\tens{X} +z_0^2-2z_0 \cos\pi\frac{2\q-1}{2P}, \quad \tens{X}=\left(e^{\tens{H}/2P}-z_0\right)/\zeta.
\label{eq:scaled-enne}
\end{equation}
Here we have introduced the shifting and scaling parameters $z_0=\left(e^{\epsilon_+/2P}+e^{\epsilon_-/2P}\right)/2$ and 
$\zeta=\left(e^{\epsilon_+/2P}-e^{\epsilon_-/2P}\right)/2$, computed in terms of the 
extremal values of the Hamiltonian spectrum, $\epsilon_-$ and $\epsilon_+$. These parameters can 
be estimated with a Lanczos procedure. Alternatively, since only a rough estimate is required,
one can also perform a test calculation on a smaller, similar system, or use a matrix norm 
as an upper bound to the spectral radius of $\tens{H}$.

The inverse $\enne{\q}^{-1}$ can be therefore approximated as a sum of Chebyshev polynomials of $\tens{X}$,
$\enne{\q}^{-1}\approx \sum_{i=0}^m c_iT_i\left(\tens{X}\right)$, where $T_i$ is the $i$-th 
Chebyshev polynomial, and the coefficients $c_i$ can
be computed, by straightforward if tedious algebra, and take the values
\begin{equation}
c_i=\frac{\delta_{i0}-2}{\zeta^2 \Im b} 
   \Im \left[\frac{\sign{\Re b}}{\sqrt{b^2-1} } \left(b-\sqrt{b^2-1} \sign \Re b\right)^i\right],
\text{ where }\quad b=\left(e^{\I\pi\frac{2\q-1}{2P}}-z_0\right)/\zeta.
\label{eq:cheb-coefficients}
\end{equation} 
An upper bound to the error due to truncation of the Chebyshev expansion can be estimated from
$\sum_{i=m+1}^{\infty}\left|c_i\right|$. This estimate leads to a complex expression, which simplifies
considerably if we assume $-\epsilon_-=\epsilon_+=\evmax$. In this case, by taking the large $P$ limit,
that the number of terms to be included in order do reach $10^{-D}$ relative accuracy on $\enne{\q}^{-1}$ reads
\begin{equation}
m_C\approx \frac{1}{2}+\frac{\evmax D \ln 10}{\pi\left(2\q-1\right)}.
\label{eq:cheb-mult}
\end{equation}
this can be used as a first guess, to be refined by explicitly summing up the $\left|c_i\right|$ as computed
from Eq.~(\ref{eq:cheb-coefficients}).

At variance with the polynomial expansion of Ref.\cite{ceri+08jcp}, the operation count is linear with $\evmax$.
This is the same scaling observed when the Fermi operator is directly expanded in Chebyshev 
polynomials\cite{baer-head97jcp}. This analogy is not surprising, since the matrix powers $T_i\left(\tens{X}\right)$
entering the expansion  are all independent of $\q$, and one could in principle obtain the whole density matrix from a single 
Chebyshev polynomial evaluation, computing the expansion coefficients by summing over all the $c_i$s.

\begin{figure}
\includegraphics[width=0.624\textwidth]{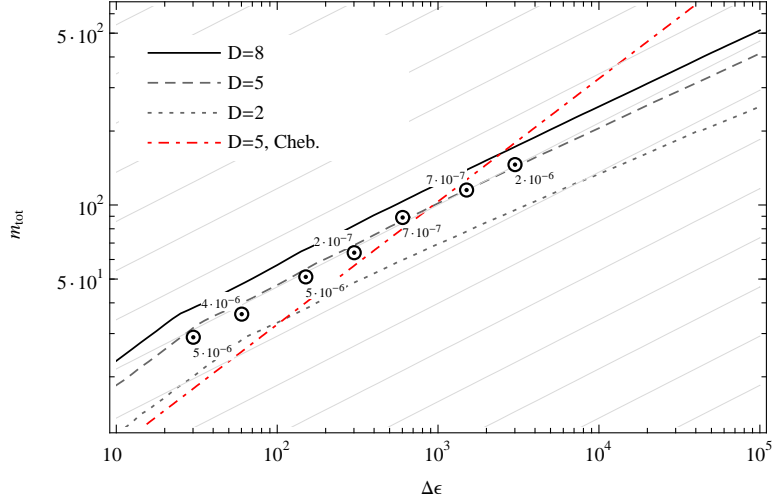}
\caption{
\label{fig:scaling}
Estimated number of required matrix-matrix multiplications needed in order to obtain $10^{-D}$ 
relative accuracy on the finite temperature density matrix, as a function of the Hamiltonian
spectum range $\evmax$. The grid lines correspond to a scaling $\propto\sqrt[3]{\evmax}$.
We also plot the operation count expected when applying the fast summation methods of 
Ref.\cite{lian+03jcp} to the standard Chebyshev polynomial expansion\cite{baer-head97jcp} 
($m_{tot}=2\sqrt{\frac{2}{3}(D-1)\evmax}$).
Data points correspond to a test performed on the {\em ab-initio} Hamiltonian for a
 $\mathrm{LiAl}$ alloy. Details can be found in~Ref.\cite{ceri+08jcp}: $\evmax$ for the system is $3.00$~a.u.,
and we have performed density matrix calculations at different temperatures, 
setting the target accuracy to $10^{-5}$. Next to each point, we also report the relative 
error on the band structure energy, as computed with our algorithm, taking as reference the value obtained by 
diagonalization at the same electronic temperature.
}
\end{figure}

However, it is more convenient to stop the summation at $\bar{\q}>1$, so as to 
reduce $m_C$, and tackle the remaining terms by iterative inversion. The practical recipe 
 is therefore to compute the coefficients $c_i\left(\bar{\q}\right)$ and 
$d_i=\sum_{\q=\bar{\q}}^P c_i\left(\q\right)$, using them to obtain $\enne{\bar{\q}}^{-1}$ and 
the contribution to the Fermi operator from the ``tail'' of matrices with $\q\ge\bar{\q}$, respectively. 
The next step is to compute the terms up to $\q=1$ by iterative inversion, a task which 
becomes much more efficient when a good initial guess for the inverse matrix is available.
Such a guess can be obtained by using a finite number of terms in the extrapolation 
\begin{equation}
\enne{\q-\delta\q}^{-1}=\frac{\enne{\q}^{-1}}{\eta}\sum_{i=0}^{\infty}
   \left[\enne{\q}^{-1}\left(1+e^{\tens{H}/P}\right)\left(1-1/\eta\right)\right]^i,\text{ with }\quad 
\eta=\cos \frac{\pi\delta\q}{p}+\sin\frac{\pi\delta\q}{p} \tan\frac{\pi\left(2\q-1\right)}{2p}.
\label{eq:newt-extrapolant}
\end{equation}
The simplest approximation $\enne{\q}^{-1}/\eta$ is already sufficient to guarantee convergence,
which is fast and almost independent of the condition number of the $\enne{\q}$'s.
The number of multiplications needed to obtain a relative accuracy of $10^{-D}$ 
on $\enne{\q}^{-1}$ starting from  $\enne{\bar{\q+1}}^{-1}/\eta$ is in fact, in the large $P$ limit,
\begin{equation}
m_N=\frac{2}{\ln 2} \ln \frac{\ln\left(1-\chi\right)-D\ln 10}{\ln\chi},\text{ where }\quad 
\chi\approx \frac{8\q}{\left(1+2\q\right)^2}.
\label{eq:newt-mult}
\end{equation}
One can use $\enne{\bar{\q}}^{-1}/\eta$, which has been computed with the Chebyshev expansion of
the tail contribution, as the initial guess for the Newton iteration which gives $\enne{\bar{\q}-1}^{-1}$.
The converged $\enne{\bar{\q-1}}^{-1}$ is in turn used to evaluate $\enne{\bar{\q}-2}^{-1}$, and so on.

The total number of matrix multiplications involved is easily obtained 
by combining Eq.~(\ref{eq:cheb-mult}) and~(\ref{eq:newt-mult}),
$m_{tot}=m_C\left(\bar{\q}\right)+\sum_{\q=1}^{\bar{\q}-1}m_N\left(\q\right)$.
By minimizing this quantity the optimal value of $\bar{\q}$ is readily found.
The Chebyshev polynomials can be computed with fast summation techniques, in which case the 
operation count for that part becomes $m_C'=3\sqrt{m_C}$. 
In Fig.~\ref{fig:scaling} we report our theoretical estimate for $m_{tot}$ as
a function of $\evmax$. The scaling is $\propto \sqrt[3]{\evmax}$, and the crossover point 
with fast-summed Chebyshev expansion is around $\evmax\approx 10^3$, which is easily reached
if electronic temperatures of a few hundred K are necessary.
In the same figure we also report the results from some test calculations performed
for the Hamiltonian of a semimetallic $\mathrm{LiAl}$ alloy, together with the 
measured relative error on the total energy, which is always one order of magnitude smaller
than the target value enforced when choosing a particular Chebyshev expansion length and  
convergence threshold for the matrix inversion.
In order to keep the scheme as simple as possible, we have not considered 
the option of fine-tuning the procedure, which could further reduce the operation count.
First, the operations count for Newton inversion~(\ref{eq:newt-mult}) turns out to be 
often overestimated, so that hand-tuning $\bar{\q}$ can save several multiplications. 
Higher-order extrapolations
are easily obtained (see Appendix~B of Ref.\cite{ceri+08jcp}), which provide
a better starting point for iterative inversion. Some preliminary results we obtained
performing molecular dynamics with forces computed on the fly with this scheme show that the number 
of multiplies needed for iterative inversion can be halved if one uses the inverse
{\em saved from the previous timestep} as the initial guess.

\section{Conclusions}
In this short paper we have introduced significant improvements to an algorithm
that tackles the problem of Fermi operator expansion by an hybrid approach based on a
combination of polynomial expansion and iterative matrix inversion, which we 
have presented in a recent paper.
With this improved implementation, only real-valued matrices are involved, leading to 
significant savings with respect to the previous, complex-valued formalism. Even more
important, the scaling of the operation count with the Hamiltonian spectrum width
is lowered from $\sqrt{\evmax}$ to $\sqrt[3]{\evmax}$, which is extremely appealing
for broad-spectrum or low electronic temperature problems.
Work in the direction of a practical, linear-scaling implementation
of these ideas is in progress\cite{melo+05cpc}. Beside the use of an efficient sparse-matrix algebra library, 
subtle issues regarding matrix truncation must be addressed, and several optimizations,
such as the extrapolation of $\emme{\q}^{-1}$ from previous timesteps discussed above,
can be used to improve the performances of the algorithm.

\section{Acknowledgements}
We would like to thank Clotilde Cucinotta and Giacomo Miceli for discussion on 
$\mathrm{LiAl}$ system, and Luca Ferraro and Simone Meloni for helping us in a preliminary
implementation of our method in a tight binding molecular dynamics code.

\bibliographystyle{aipproc}

\begin{thebibliography}{17}
\expandafter\ifx\csname natexlab\endcsname\relax\def\natexlab#1{#1}\fi
\providecommand{\enquote}[1]{``#1''}
\expandafter\ifx\csname url\endcsname\relax
  \def\url#1{\texttt{#1}}\fi
\expandafter\ifx\csname urlprefix\endcsname\relax\def\urlprefix{URL }\fi
\providecommand{\eprint}[2][]{\url{#2}}

\bibitem[Kohn(1996)]{kohn96prl}
W.~Kohn, \emph{Phys. Rev. Lett.} \textbf{76}, 3168--3171 (1996).

\bibitem[Goedecker(1999)]{goed99rmp}
S.~Goedecker, \emph{Rev. Mod. Phys.} \textbf{71}, 1085--1123 (1999).

\bibitem[Yang(1991)]{yang91prl}
W.~Yang, \emph{Phys. Rev. Lett.} \textbf{66}, 1438--1441 (1991).

\bibitem[Galli and Parrinello(1992)]{gall-parr92prl}
G.~Galli, and M.~Parrinello, \emph{Phys. Rev. Lett.} \textbf{69},
  3547--3550 (1992).

\bibitem[Li et~al.(1993)]{li+93prb}
X.~P. Li, R.~W. Nunes, and D.~Vanderbilt, \emph{Phys. Rev. {\bf B}} \textbf{47},
  10891--10894 (1993).

\bibitem[Baroni and Giannozzi(1992)]{baro-gian92epl}
S.~Baroni, and P.~Giannozzi, \emph{Europhys. Lett.} \textbf{17}, 547
  (1992).

\bibitem[Palser and Manolopoulos(1998)]{pals-mano98prb}
A.~H.~R. Palser, and D.~E. Manolopoulos, \emph{Phys. Rev. {\bf B}} \textbf{58},
  12704--12711 (1998).

\bibitem[Goedecker and Colombo(1994)]{goed-colo94prl}
S.~Goedecker, and L.~Colombo, \emph{Phys. Rev. Lett.} \textbf{73},
  122--125 (1994).

\bibitem[Goedecker and Teter(1995)]{goed-tete95prb}
S.~Goedecker, and M.~Teter, \emph{Phys. Rev. B} \textbf{51}, 9455--9464
  (1995).

\bibitem[Ceriotti et~al.(2008)]{ceri+08jcp}
M.~Ceriotti, T.~D. K{\"u}hne, and M.~Parrinello, \emph{J. Chem.
  Phys.} \textbf{129}, 024707 (2008).

\bibitem[Alavi et~al.(1994)]{alav+94prl}
A.~Alavi, J.~Kohanoff, M.~Parrinello, and D.~Frenkel, \emph{Phys. Rev.
  Lett.} \textbf{73}, 2599 (1994).

\bibitem[Krajewski and Parrinello(2005)]{kraj-parr05prb}
F.~R. Krajewski, and M.~Parrinello, \emph{Phys. Rev. {\bf B}} \textbf{71},
  233105 (2005).

\bibitem[Pan and Reif(1985)]{pan-reif85proc}
V.~Pan, and J.~Reif, \enquote{Efficient parallel solution of linear systems,}
  in \emph{STOC {'}85: Proceedings of the seventeenth annual ACM symposium on
  Theory of computing}, ACM Press, New York, NY, USA, 1985, p. 143, ISBN
  0-89791-151-2.

\bibitem[Baer and Head-Gordon(1997)]{baer-head97jcp}
R.~Baer, and M.~Head-Gordon, \emph{J. Chem. Phys.} \textbf{107}, 10003--10013
  (1997).

\bibitem[Van~Loan(1979)]{vanl79ieee}
C.~Van~Loan, \emph{IEEE Trans. on Automatic Control} \textbf{24},
  320--321 (1979).

\bibitem[Liang et~al.(2003)]{lian+03jcp}
W.~Z. Liang, C.~Saravanan, Y.~Shao, R.~Baer, A.~T. Bell, and M.~Head-Gordon,
  \emph{J. Chem. Phys.} \textbf{119}, 4117 (2003).

\bibitem[Meloni et~al.(2005)]{melo+05cpc}
S.~Meloni, M.~Rosati, A.~Federico, L.~Ferraro, A.~Mattoni, and L.~Colombo,
  \emph{Comp. Phys. Comm.} \textbf{169}, 462--466 (2005).

\end{thebibliography}
\end{document}